\documentclass{acm_proc_article-sp}
\usepackage{url}
\usepackage{graphicx}

\newcommand{\figwidth}{0.87\columnwidth}

\begin{document}

\title{Predicting the popularity of online content}

\numberofauthors{2}

\author{
\alignauthor
Gabor Szabo\\
\affaddr{Social Computing Lab}\\
\affaddr{HP Labs}\\
\affaddr{Palo Alto, CA}\\
\email{gabors@hp.com}
\alignauthor
Bernardo A. Huberman\\
\affaddr{Social Computing Lab}\\
\affaddr{HP Labs}\\
\affaddr{Palo Alto, CA}\\
\email{bernardo.huberman@hp.com}
}

\date{\today}

\newcommand{\youtube}{Youtube}
\newcommand{\digg}{Digg}
\newcommand{\slashdot}{Slashdot}

\maketitle

\begin{abstract}
We present a method for accurately predicting the long time popularity
of online content from early measurements of user's access. Using two
content sharing portals, \youtube{} and \digg{}, we show that by
modeling the accrual of views and votes on content offered by these
services we can predict the long-term dynamics of individual
submissions from initial data. In the case of \digg{}, measuring
access to given stories during the first two hours allows us to
forecast their popularity 30 days ahead with remarkable accuracy,
while downloads of \youtube{} videos need to be followed for 10 days
to attain the same performance. The differing time scales of the
predictions are shown to be due to differences in how content is
consumed on the two portals: \digg{} stories quickly become outdated,
while \youtube{} videos are still found long after they are initially
submitted to the portal. We show that predictions are more accurate
for submissions for which attention decays quickly, whereas
predictions for evergreen content will be prone to larger errors.
\end{abstract}


\section{Introduction}

The ubiquity and inexpensiveness of Web 2.0 services have transformed
the landscape of how content is produced and consumed online. Thanks
to the web, it is possible for content producers to reach out to
audiences with sizes that are inconceivable using conventional
channels. Examples of the services that have made this exchange
between producers and consumers possible on a global scale include
video, photo, and music sharing, weblogs and wikis, social bookmarking
sites, collaborative portals, and news aggregators where content is
submitted, perused, and often rated and discussed by the user
community. At the same time, the dwindling cost of producing and
sharing content has made the online publication space a highly
competitive domain for authors.

The ease with which content can now be produced brings to the center
the problem of the attention that can be devoted to it. Recently, it
has been shown that attention~\cite{citeulike:2481055} is allocated in
a rather asymmetric way, with most content getting some views and
downloads, whereas only a few receive the bulk of the attention. While
it is possible to predict the distribution in attention over many
items, so far it has been hard to predict the amount that would be
devoted over time to given ones. This is the problem we solve in this
paper.

Most often portals rank and categorize content based on its quality
and appeal to users. This is especially true of aggregators where the
``wisdom of the crowd'' is used to provide collaborative filtering
facilities to select and order submissions that are favored by many.
One such well-known portal is \digg{}, where users submit links and
short descriptions to content that they have found on the Web, and
others vote on them if they find the submission interesting. The
articles collecting the most votes are then exhibited on premiere
sections across the site, such as the ``recently popular submissions''
(the main page), and ``most popular of the day/week/month/year''. This
results in a positive feedback mechanism that leads to a ``rich get
richer'' type of vote accrual for the very popular items, although it
is also clear that this pertains to only a very small fraction of the
submissions.

As a parallel to \digg{}, where content is not produced by the
submitters themselves but only linked to it, we study \youtube{}, one
of the first video sharing portals that lets users upload, describe,
and tag their own videos. Viewers can watch, reply to, and leave
comments on them. The extent of the online ecosystem that has
developed around the videos on \youtube{} is impressive by any
standards, and videos that draw a lot of viewers are prominently
exposed on the site, similarly to \digg{} stories.

The paper is organized as follows. In Section~\ref{section:data} we
describe how we collected access data on submissions on \youtube{} and
\digg{}. Section~\ref{section:daily cycles} shows how daily or weekly
fluctuations can be observed in \digg{}, together with presenting a
simple method to eliminate them for the sake of more accurate
predictions. In Section~\ref{section:predictions} we discuss the
models used to describe content popularity and how prediction accuracy
depends on their choice. Here we will also point out that the expected
growth in popularity of videos on \youtube{} is markedly different
from when compared to \digg{}, and further study the reasons for this
in Section~\ref{section:saturation}. In
Section~\ref{section:conclusions} we conclude and cite relevant works
to this study.

\section{Sources of data}
\label{section:data}

The formulation of the prediction models relies heavily on observed
characteristics of our experimental data, which we describe in this
section. The organization of \youtube{} and \digg{} is conceptually
similar to each other, so we can also employ a similar framework to
study content popularity after the data has been normalized. To
simplify the terminology, by \emph{popularity} in the following we
will refer to the number of views that a video receives on \youtube{},
and to the number of votes (diggs) that a story collects on \digg{},
respectively.

\subsection{Youtube}

\youtube{} is the pinnacle of user-created video sharing portals on
the Web, with 65,000 new videos uploaded and 100 million downloaded on
a daily basis, implying that that 60\% of all online videos are
watched through the portal~\cite{1298310}. \youtube{} is also the
third most frequently accessed site on the Internet based on traffic
rank~\cite{1298310,cheng:2007,alexa}. We started collecting view count
time series on 7,146 selected videos daily, beginning April 21, 2008,
on videos that appeared in the ``recently added'' section of the
portal on this day. Apart from the list of most recently added videos,
the web site also offers listings based on different selection
criteria, such as ``featured'', ``most discussed'', and ``most
viewed'' lists, among others. We chose the most recently uploaded list
to have an unbiased sample of all videos submitted to the site in the
sampling period, not only the most popular ones, and also so that we
can have a complete history of the view counts for each video during
their lifetime. The \youtube{} application programming
interface~\cite{youtube_api} gives programmatic access to several of a
video's statistics, the view count at a given time being one of them.
However, due to the fact that the view count field of a video does not
appear to be updated more often than once a day by \youtube{}, it is
only possible to have a good approximation for the number of views
daily. Within a day, however, the API does indicate when the view
count was recorded. It is worth noting that while the overwhelming
majority of video views is initiated from the \youtube{} website
itself, videos may be linked from external sources as well (about half
of all videos are thought to be linked externally, but also that only
about 3\% of the views are coming from these
links~\cite{citeulike:2506755}).

In Section~\ref{section:predictions}, we compare the view counts of
videos at given times after their upload. Since in most cases we only
have information on the view counts once a day, we use linear
interpolation between the nearest measurement points around the time
of interest to approximate the view count at the given time.

\subsection{Digg}

\digg{} is a Web 2.0 service where registered users can submit links
and short descriptions to news, images, or videos they have found
interesting on the Web, and which they think should hold interest for
the greater general audience, too ($90.5\%$ of all uploads were links
to news, $9.2\%$ to videos, and only $0.3\%$ to images). Submitted
content will be placed on the site in a so-called ``upcoming''
section, which is one click away from the main page of the site. Links
to content are provided together with surrogates for the submission (a
short description in the case of news, and a thumbnail image for
images and videos), which is intended to entice readers to peruse the
content. The main purpose of \digg{} is to act as a massive
collaborative filtering tool to select and show the most popular
content, and thus registered users can \emph{digg} submissions they
found interesting. This serves to increase the digg count of the
submission by one, and submissions that get substantially enough diggs
in a relatively short time in the upcoming section will be presented
on the front page of \digg{}, or using its terminology, they will be
\emph{promoted} to the front page. Someone's submission being promoted
is a considerable source of pride in the \digg{} community, and is a
main motivator for returning submitters. The exact algorithm for
promotion is not made public to thwart gaming, but is thought to give
preference to upcoming submissions that accumulate diggs quickly
enough from diverse neighborhoods of the \digg{} social
network~\cite{Lerman07ic}. The social networking feature offered by
\digg{} is extremely important, through which users may place watch
lists on another user by becoming their ``fans''. Fans will be shown
updates on which submissions users dugg who they are fans of, and thus
the social network will play a major role in making upcoming
submissions more visible. Very importantly, in this paper we also only
consider stories that were promoted to the front page, given that we
are interested in submissions' popularity among the general user base
rather than in niche social networks.

We used the \digg{} API~\cite{digg_api} to retrieve all the diggs made
by registered users between July 1, 2007, and December 18, 2007. This
data set comprises of about 60 million diggs by 850 thousand users in
total, cast on approximately 2.7 million submissions (this number
includes all past submissions also that received any digg). The number
of submissions in this period was 1,321,903, of which 94,005 (7.1\%)
became promoted to the front page.

\section{Daily cycles}
\label{section:daily cycles}

\begin{figure}
\begin{center}
\includegraphics[width=\figwidth]{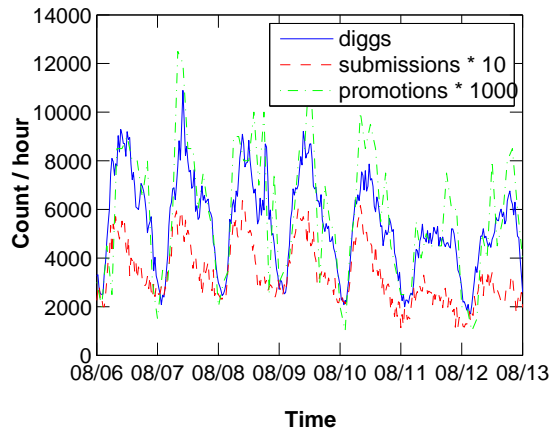}
\end{center}
\caption{Daily and weekly cycles in the hourly rates of digging
  activity, story submissions, and story promotions, respectively. To
  match the different scales the rates for submissions is multiplied
  by 10, that of the promotions is multiplied by 1000. The horizontal
  axis represents one week from August 6, 2007 (Monday) through Aug
  12, 2007 (Sunday). The tick marks represent midnight of the
  respective day, Pacific Standard Time.}
\label{figure:daily_activities}
\end{figure}

In this section we examine the daily and weekly activity variations in
user activity. Figure~\ref{figure:daily_activities} shows the hourly
rates of digging and story submitting of users, and of upcoming story
promotions by \digg{}, as a function of time for one week, starting
August 6, 2007. The difference in the rates may be as much as
threefold, and weekends also show lesser activity. Similarly,
Fig.~\ref{figure:daily_activities} also showcases weekly variations,
where weekdays appear about 50\% more active than weekends. It is also
reasonable to assume that besides daily and weekly cycles, there are
seasonal variations as well. It may also be concluded that \digg{}
users are mostly located in the UTC-5 to UTC-8 time zones, and since
the official language of \digg{} is English, \digg{} users are mostly
from North America.

\begin{figure}
\begin{center}
\includegraphics[width=\figwidth]{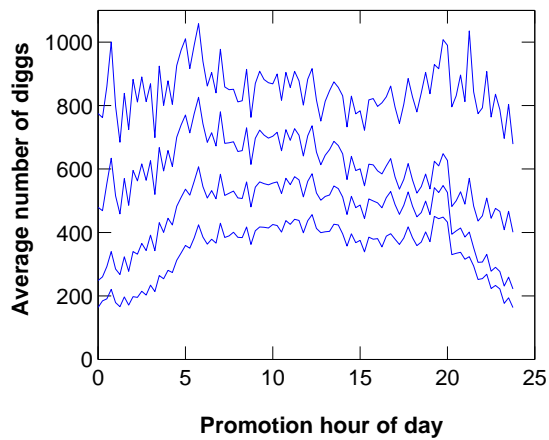}
\end{center}
\caption{The average number of diggs that stories get after a certain
  time, shown as a function of the hour that the story was promoted at
  (PST). Curves from bottom to top correspond to measurements made 2,
  4, 8, and 24 hours after promotion, respectively.}
\label{figure:digg_count_depending_on_promotion_time_of_day}
\end{figure}

Depending on the time of day when a submission is made to the portal,
stories will differ greatly on the number of initial diggs that they
get, as
Fig.~\ref{figure:digg_count_depending_on_promotion_time_of_day}
illustrates. As can be expected, stories submitted at less active
periods of the day will accrue less diggs in the first few hours
initially than stories submitted during peak times. This is a natural
consequence of suppressed digging activity during the nightly hours,
but may initially penalize interesting stories that will ultimately
become popular. In other words, based on observations made only after
a few hours after a story has been promoted, we may misinterpret a
story's relative interestingness, if we do not correct for the
variation in daily activity cycles. For instance, a story that gets
promoted at 12pm will on average get approximately 400 diggs in the
first 2 hours, while it will only get 200 diggs if it is promoted at
midnight.

Since the digging activity varies by time, we introduce the notion of
\emph{digg time}, where we measure time not by wall time (seconds),
but by the number of all diggs that users cast on promoted stories. We
choose to count diggs only on promoted stories only because this is
the section of the portal that we focus on stories from, and most
diggs ($72\%$) are going to promoted stories anyway. The average
number of diggs arriving to promoted stories per hour is 5,478 when
calculated over the full data collection period, thus we define one
digg hour as the time it takes for so many new diggs to be cast. As
seen earlier, during the night this will take about three times longer
than during the active daily periods. This transformation allows us to
mitigate the dependence of submission popularity on the time of day
when it was submitted. When we refer to the age of a submission in
digg hours at a given time $t$, we measure how many diggs were
received in the system between $t$ and the submission of the story,
and divide by 5,478. A further reason to use digg time instead of
absolute time will be given in Section~\ref{section:correlations}.

\section{Predictions}
\label{section:predictions}

In this section we show that if we perform a logarithmic
transformation on the popularities of submissions, the transformed
variables exhibit strong correlations between early and later times,
and on this scale the random fluctuations can be expressed as an
additive noise term. We use this fact to model and predict the future
popularity of individual content, and measure the performance of the
predictions.

In the following, we call \emph{reference time} $t_r$ the time when we
intend to predict the popularity of a submission whose age with
respect to the upload (promotion) time is $t_r$. By \emph{indicator
  time} $t_i$ we refer to when in the life cycle of the submission we
perform the prediction, or in other words how long we can observe the
submission history in order to extrapolate; $t_i < t_r$.

\subsection{Correlations between early and later times}
\label{section:correlations}

\begin{figure}
\begin{center}
\includegraphics[width=\figwidth]{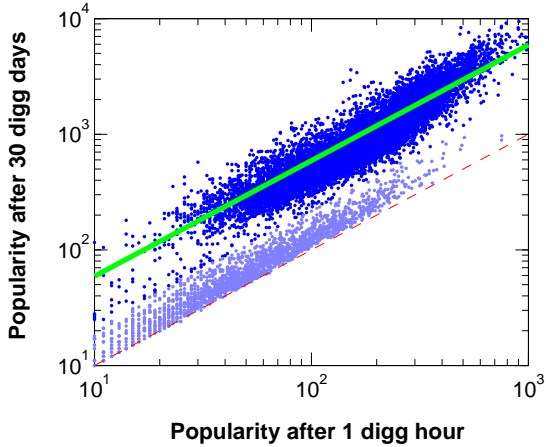}
\end{center}
\caption{The correlation between digg counts on the 17,097 promoted
  stories in the dataset that are older than 30 days. A k-means
  clustering separates 89\% of the stories into the upper cluster,
  while the rest of the stories is shown in lighter color. The bold
  guide line indicates a linear fit with slope 1 on the upper cluster,
  with a prefactor of 5.92 (the Pearson correlation coefficient is
  0.90). The dashed line marks the $y = x$ line below which no stories
  can fall.}
\label{figure:digg-correlations}
\end{figure}

\begin{figure}
\begin{center}
\includegraphics[width=\figwidth]{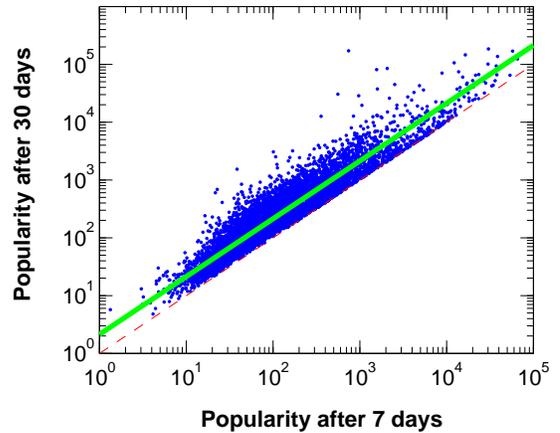}
\end{center}
\caption{The popularities of videos shown at the 30th day after
  upload, versus their popularity after 7 days. The bold solid line
  with gradient 1 has been fit to the data, with correlation
  coefficient $R = 0.77$ and prefactor $2.13$.}
\label{figure:youtube-correlations}
\end{figure}

We first consider the question whether the popularity of submissions
early on is any predictor of their popularity at a later stage, and if
so, what the relationship is. For this, we first plot the popularity
counts for submissions at the reference time $t_r = 30$ days both for
\digg{} (Fig.~\ref{figure:digg-correlations}) and \youtube{}
(Fig.~\ref{figure:youtube-correlations}), versus the popularities
measured at the indicator times $t_i = 1$ digg hour, and $t_i = 7$
days for the two portals, respectively. We choose to measure the
popularity of \youtube{} videos at the end of the 7th day so that the
view counts at this time are in the $10^1$--$10^4$ range, and
similarly for \digg{} in this measurement. We logarithmically rescale
the horizontal and vertical axes in the figures due to the large
variances present among the popularities of different submissions
(notice that they span three decades).

Observing the \digg{} data, one notices that the popularity of about
11\% of stories (indicated by lighter color in
Fig.~\ref{figure:digg-correlations}) grows much slower than that of
the majority of submissions: by the end of the first hour of their
lifetime, they have received most of the diggs that they will ever
get. The separation of the two clusters is perceivable until
approximately the 7th digg hour, after which the separation vanishes
due to fact that by that time the digg counts of stories mostly
saturate to their respective maximum values (skip to
Fig.~\ref{figure:both-average-normalized-pop} for the average growth
of \digg{} article popularities). While there is no obvious reason for
the presence of clustering, we assume that it arises when the
promotion algorithm of \digg{} misjudges the expected future
popularity of stories, and promotes stories from the upcoming phase
that will not maintain a sustained attention from the users. Users
thus lose interest in them much sooner than in stories in the upper
cluster. We used k-means clustering with $k=2$ and cosine distance
measure to separate the two clusters as shown in
Fig.~\ref{figure:digg-correlations} up to the 7th digg hour (after
which the clusters are not separable), and we exclusively use the
upper cluster for the calculations in the following.

As a second step, to quantify the strength of the correlations
apparent in Figs.~\ref{figure:digg-correlations} and
\ref{figure:youtube-correlations}, we measured the Pearson correlation
coefficients between the popularities at different indicator times and
the reference time. The reference time is always chosen $t_r = 30$
days (or digg days for \digg{}) as previously, and the indicator time
is varied between $0$ and $t_r$.

\begin{figure}
\begin{center}
\includegraphics[width=\figwidth]{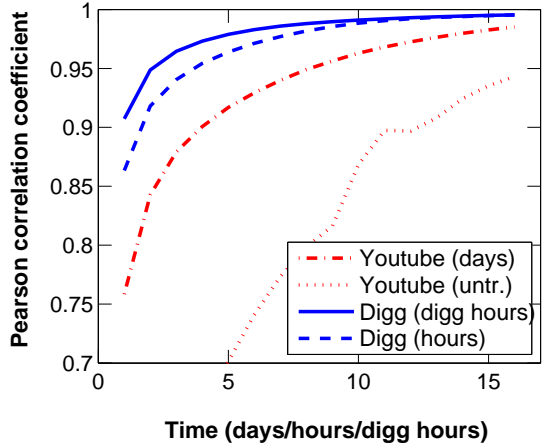}
\end{center}
\caption{The Pearson correlation coefficients between the logarithms
  of the popularities of submissions measured at different times: at
  the time indicated by the horizontal axis, and on the 30th day. For
  \youtube{}, the x-axis is in days. For \digg{}, it is in hours for
  the dashed line, and digg hours for the solid line (stronger
  correlation). For comparison, the dotted line shows the correlation
  coefficients for the untransformed (non-logarithmic) popularities in
  \youtube{}.}
\label{figure:both-correlation_coefficients}
\end{figure}

\emph{\youtube{}}. Fig.~\ref{figure:both-correlation_coefficients}
shows the Pearson correlation coefficients between the logarithmically
transformed popularities, and for comparison also the correlations
between the untransformed variables. The PCC is $0.92$ after about 5
days; however, the untransformed scale shows weaker linear dependence,
at 5 days the PCC is only $0.7$, and it consistently stays below the
PCC of the logarithmically transformed scale.

\emph{\digg{}}. Also in
Fig.~\ref{figure:both-correlation_coefficients}, we plot the PCCs of
the log-transformed popularities between the indicator times and the
reference time. It is already $0.98$ after the 5th digg hour, and it
is as strong as $0.993$ after the 12th. We also argue here that by
measuring submission age as digg time leads to stronger correlations:
the figure shows the PCC as well for the case when the story age is
measured as absolute time (dashed line, 17,222 stories), and it is
always less than the PCCs taken with digg hours (solid line, 17,097
stories) up to approximately the 12th hour. This is understandable
since this is the time scale of the strongest daily variations (cf.
Fig.~\ref{figure:daily_activities}). We do not show the untransformed
scale PCC for \digg{} submissions measured in digg hours, since it
approximately traces the dashed line in the figure, thus also
indicating a weaker correlation than the solid line.

\subsection{The evolution of submission popularity}

The strong linear correlation found between the indicator and
reference times of the logarithmically transformed submission
popularities suggests that the more popular submissions are in the
beginning, the more they will be also later on, and the connection can
be described by a linear model:
\begin{eqnarray}
\label{equation:linear_product}
\ln N_s(t_2) & = & \ln \left[ r(t_1, t_2) N_s(t_1) \right] + \xi_s(t_1,
t_2) \\
\nonumber
& = & \ln r(t_1, t_2) + \ln N_s(t_1) + \xi_s(t_1, t_2),
\end{eqnarray}
where $N_s(t)$ is the popularity of submission $s$ at time $t$ (in the
case of \digg{}, time is naturally measured by digg time), and $t_1$
and $t_2$ are two arbitrarily chosen points in time, $t_2 > t_1$.
$r(t_1, t_2)$ accounts for the linear relationship found between the
log-transformed popularities at different times, and it is independent
of $s$. $\xi_s$ is a noise term drawn from a given distribution with
mean 0 that describes the randomness observed in the data. It is
important to note that the noise term is additive on the log-scale of
popularities, justified by the fact that the strongest correlations
were found on this transformed scale. Considering
Figures~\ref{figure:digg-correlations} and
\ref{figure:youtube-correlations}, the popularities at $t_2 = t_r$
also appear to be evenly distributed around the linear fit (with
taking only the upper cluster in Fig.~\ref{figure:digg-correlations}
and considering the natural cutoff $y = x$ in
Fig.~\ref{figure:youtube-correlations}).

We will now show that the variations of the log-popularities around
the expected average are distributed approximately normally with an
additive noise. To this end we performed linear regression on the
logarithmicalyy transformed data points shown in
Figs.~\ref{figure:digg-correlations} and
\ref{figure:youtube-correlations}, respectively, fixing the slope of
the linear regression function to $1$ in accordance with
Eq.~(\ref{equation:linear_product}). The intercept of the linear fit
corresponds to $\ln r(t_i, t_r)$ above ($t_i = 7$ days/$1$ digg hour,
$t_r = 30$ days), and $\xi_s(t_i, t_r)$ are given by the residuals of
the variables with respect to the best fit.

\begin{figure}
\begin{center}
\includegraphics[width=\figwidth]{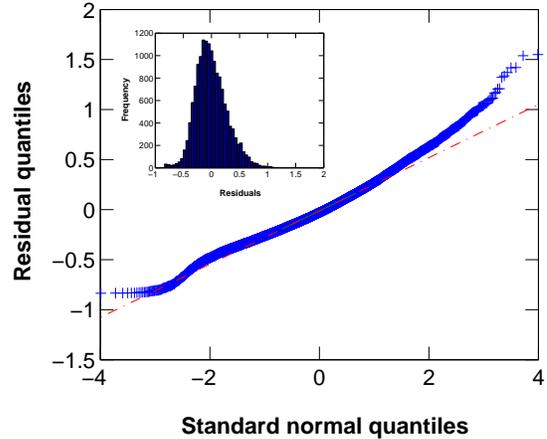}
\end{center}
\caption{The quantile-quantile plot of the residuals of the linear fit
  of Fig.~\ref{figure:digg-correlations} to the logarithms of \digg{}
  story popularities, as described in the text. The inset shows the
  frequency distribution of the residuals.}
\label{figure:digg-normal_quantiles}
\end{figure}

\begin{figure}
\begin{center}
\includegraphics[width=\figwidth]{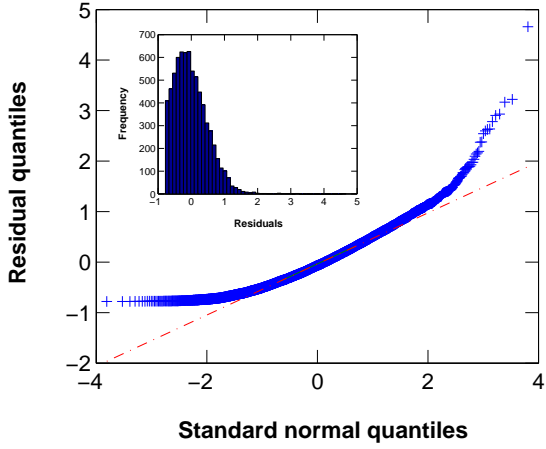}
\end{center}
\caption{The quantile-quantile plot of the residuals of the linear fit
  of Fig.~\ref{figure:youtube-correlations} for \youtube{}.}
\label{figure:youtube-normal_quantiles}
\end{figure}

We tested the normality of the residuals by plotting the quantiles of
their empirical distributions versus the quantiles of the theoretical
(normal) distributions in Figs.~\ref{figure:digg-normal_quantiles}
(\digg{}) and \ref{figure:youtube-normal_quantiles} (\youtube{}). The
residuals show a reasonable match with normal distributions, although
we observe in the quantile-quantile plots that the measured
distributions of the residuals are slightly right-skewed, which means
that content with very high popularity values is overrepresented in
comparison to less popular content. This is understandable if we
consider that a small fraction of the submissions ends up on ``most
popular'' and ``top'' pages of both portals. These are the submissions
that are deemed most requested by the portals, and are shown to the
users as those that others found most interesting. They stay on
frequented and very visible parts of the portals, and are naturally
attract further diggs/views. In the case of \youtube{}, one can see
that content popularity at the 30th day versus the 7th day as shown in
Fig.~\ref{figure:youtube-correlations} is bounded from below, due to
the fact the view counts can only grow, and thus the distribution of
residuals is also truncated in
Fig.~\ref{figure:youtube-normal_quantiles}. We also note that the
Jarque-Bera and Lilliefors tests reject residual normality at the 5\%
significance level for both systems, although the residuals appear to
be distributed reasonably close to Gaussians. Moreover, to see whether
the homoscedasticity of the residuals holds that is necessary for the
linear regression [their variance being independent of $N_s(t_i)$], we
checked the means and variances of the residuals as a function of
$N_c(t_i)$ by subdividing the popularity values into 50 bins, with the
result that both the mean and variance are independent of $N_c(t_i)$.

A further justification for the model of
Eq.~(\ref{equation:linear_product}) is given in the following. It has
been shown that the popularity distribution of \digg{} stories of a
given age follows a lognormal distribution~\cite{citeulike:2481055}
that is the result of a growth mechanism with multiplicative noise,
and can be described as
\begin{equation}
\ln N_s(t_2) = \ln N_s(t_1) + \sum_{\tau = t_1}^{t_2} \eta(\tau),
\end{equation}
where $\eta(\cdot)$ denotes independent values drawn from a fixed
probability distribution, and time is measured in discrete steps. If
the difference between $t_1$ and $t_2$ is large enough, the
distribution of the sum of $\eta(\tau)$'s will approximate a normal
distribution, according to the central limit theorem. We can thus map
the mean of the sum of $\eta(\tau)$'s to $\ln r(t_1, t_2)$ in
Eq.~(\ref{equation:linear_product}), and find that the two
descriptions are equivalent characterizations of the same lognormal
growth process.

\subsection{Prediction models}

We present three models to predict an individual submission's
popularity at a future time $t_r$. The performance of the predictions
is measured on the test sets by defining error functions that yield a
measure of deviation of the predictions from the observed popularities
at $t_r$, and together with the models we discuss what error measure
they are expected to minimize. One model that minimizes a given error
function may fare worse for another error measure.

The first prediction model closely parallels the experimental
observations shown in the previous section. In the second, we consider
a common error measure and formulate the model so that it is optimal
with respect to this error function. Lastly, the third prediction
method is presented as comparison and one that has been used in
previous works as an ``intuitive'' way of modeling popularity
growth~\cite{1318432}. Below, we use the $\hat x$ notation to refer to
the predicted value of $x$ at $t_r$.

\subsubsection{LN model: linear regression on a logarithmic scale;
  least-squares absolute error}

The linear relationship found for the logarithmically transformed
popularities and described by Eq.~(\ref{equation:linear_product})
above suggests that given the popularity of a submission at a given
time, a good estimate we can give for a later time is determined by
the ordinary least squares estimate, and it is the best estimate that
minimizes the sum of the squared residuals (a consequence of the
linear regression with the maximum likelihood method). However, the
linear regression assumes normally distributed residuals and the
lognormal model gives rise to additive Gaussian noise only if the
logarithms of the popularities are considered, and thus the overall
error that is minimized by the linear regression on this scale is
\begin{equation}
\mathrm{LSE}^* = \sum_c r_c^2 = \sum_c \left[ \hat \ln N_c(t_i, t_r) -
  \ln N_c(t_r) \right] ^ 2,
\label{equation:LSE-log}
\end{equation}
where $\hat \ln N_c(t_i, t_r)$ is the prediction for $\ln N_c(t_r)$,
and is calculated as $\hat \ln N_c(t_i, t_r) = \beta_0(t_i) + \ln
N_c(t_i)$ and $\beta_0$ is yielded by the maximum likelihood parameter
estimator for the intercept of the linear regression with slope $1$.
The sum in Eq.~(\ref{equation:LSE-log}) goes over all content in the
training set when estimating the parameters, and the test set when
estimating the error. We, on the other hand, are in practice
interested in the error on the linear scale,
\begin{equation}
\mathrm{LSE} = \sum_c \left[ \hat N_c(t_i, t_r) - N_c(t_r) \right] ^ 2.
\end{equation}
The residuals, while distributed normally on the logarithmic scale,
will not have this property on the untransformed scale, and an
inconsistent estimate would result if we used $\exp \left[ \hat \ln
  N_c(t_i, t_r) \right]$ as a predictor on the natural (original)
scale of popularities~\cite{duan:1983}. However, fitting least squares
regression models to transformed data has been extensively
investigated (see Refs.
\cite{duan:1983,RePEc:spr:empeco:v:18:y:1993:i:2:p:307-19,RePEc:ier:iecrev:v:33:y:1992:i:4:p:935-55}),
and in case the transformation of the dependent variable is
logarithmic, the best untransformed scale estimate is
\begin{equation}
\hat N_s(t_i, t_r) = \exp\left[ \ln N_s(t_i) + \beta_0(t_i) + \sigma_0
  ^ 2 / 2 \right].
\label{equation:inverse_box-cox}
\end{equation}
Here $\sigma_0 ^ 2 = \mathrm{var}(r_c)$, the consistent estimate
for the variance of the residuals on the logarithmic scale. Thus the
procedure to estimate the expected popularity of a given submission
$s$ at time $t_r$ from measurements at time $t_i$, we first determine
the regression coefficient $\beta_0(t_i)$ and the variance of the
residuals $\sigma_0^2$ from the training set, and apply
Eq.~(\ref{equation:inverse_box-cox}) to obtain the expectation on the
original scale, using the popularity $N_s(t_i)$ measured for $s$ at
$t_i$.

\subsubsection{CS model: constant scaling model; relative squared
  error}

In this section we first define the error function that we wish to
minimize, and then present a linear estimator for the predictions.

The relative squared error that we use here takes the form of
\begin{equation}
\mathrm{RSE} = \sum_c \left[ \frac{\hat N_c(t_i, t_r) -
    N_c(t_r)}{N_c(t_r)} \right]^2 = \sum_c \left[ \frac{\hat N_c(t_i,
    t_r)}{N_c(t_r)} - 1 \right]^2.
\label{equation:relative_squared_error}
\end{equation}
This is similar to the commonly used relative standard error
\begin{equation}
\left| \frac{\hat N_c(t_i, t_r) - N_c(t_r)}{N_c(t_r)} \right|,
\end{equation}
except that the absolute value of the relative difference is replaced
by a square.

The linear correspondence found between the logarithms of the
popularities up to a normally distributed noise term suggests that the
future expected value $\hat N_s(t_i, t_r)$ for submission $s$ can be
expressed as
\begin{equation}
\hat N_s(t_i, t_r) = \alpha(t_i, t_r) N_s(t_i).
\label{equation:constant_growth_factor}
\end{equation}
$\alpha(t_i, t_r)$ is independent of the particular submission $s$,
and only depends on the indicator and reference times. The value that
$\alpha(t_i, t_r)$ takes, however, will be contingent on what the
error function is, so that the optimal value of $\alpha$ minimizes
this. We will minimize RSE on the training set if and only if
\begin{equation}
0 = \frac{\partial \mathrm{RSE}}{\partial \alpha(t_i, t_r)} = 2 \sum_c
\left[ \frac{N_c(t_i)}{N_c(t_r)} \alpha(t_i, t_r) - 1 \right]
\frac{N_c(t_i)}{N_c(t_r)}.
\end{equation}
Expressing $\alpha(t_i, t_r)$ from above,
\begin{equation}
\alpha(t_i, t_r) = \frac{\sum_c \frac{N_c(t_i)}{N_c(t_r)}}{\sum_c
  \left[ \frac{N_c(t_i)}{N_c(t_r)} \right]^2}.
\end{equation}
The value of $\alpha(t_i, t_r)$ can be calculated from the training
data for any $t_i$, and further, the prediction for any new submission
may be made knowing its age using this value from the training set,
together with Eq.~(\ref{equation:constant_growth_factor}). If we
verified the error on the training set itself, it is guaranteed that
RSE is minimized under the model assumptions of linear scaling.

\subsubsection{GP model: growth profile model}

For comparison, we consider a third description for predicting future
content popularity, which is based on average growth profiles devised
from the training set~\cite{1318432}. This assumes in essence that the
growth of a submission's popularity in time follows a uniform accrual
curve, which is appropriately rescaled to account for the differences
between submission interestingnesses. The growth profile is calculated
on the training set as the average of the relative popularities of the
submissions of a given age $t_i$, as normalized by the final
popularity at the reference, $t_r$:
\begin{equation}
P(t_0, t_1) = \left\langle \frac{N_c(t_0)}{N_c(t_1)} \right\rangle_c,
\label{equation:norm-popularity}
\end{equation}
where $\langle \cdot \rangle_c$ takes the mean of its argument over
all content in the training set. We assume that the rescaled growth
profile approximates the observed popularities well over the whole
time axis with an affine transformation, and thus at $t_i$ the
rescaling factor $\Pi_s$ is given by $N_s(t_i) = \Pi_s(t_i, t_r)
P(t_i, t_r)$. The prediction for $t_r$ consists of using $\Pi_s(t_i,
t_r)$ to calculate the future popularity,
\begin{equation}
\hat N_s(t_r) = \Pi_s(t_i, t_r) P(t_r, t_r) = \Pi_s(t_i, t_r) =
\frac{N_s(t_i)}{P(t_i, t_r)}.
\end{equation}
The growth profiles for \youtube{} and \digg{} were measured and shown
in Fig.~\ref{figure:both-average-normalized-pop}.

\subsection{Prediction performance}

The performance of the prediction methods will be assessed in this
section, using two error functions that are analogous to LSE and RSE,
respectively.

We subdivided the submission time series data into a training set and
a test set, on which we benchmarked the different prediction schemes.
For \digg{}, we took all stories that were submitted during the first
half of the data collection period as the training set, and the second
half was considered as the test set. On the other hand, the 7,146
\youtube{} videos that we followed were submitted around the same
time, so instead we randomly selected 50\% of these videos as training
and the other half as test. The number of submissions that the
training and test sets contain are summarized in
Table~\ref{table:sets}. The parameters defined in the prediction
models were found through linear regression ($\beta_0$ and
$\sigma_0^2$) and sample averaging ($\alpha$ and $P$), respectively.

\begin{table}
\begin{center}
\begin{tabular*}{0.9\columnwidth}{@{\extracolsep{\fill}} l|l|l}
 & Training set & Test set \\ \hline
\digg{} & 10825 stories & 6272 stories \\
 & (7/1/07--9/18/07) & (9/18/07--12/6/07) \\ \hline
\youtube{} & 3573 videos & 3573 videos \\
 & randomly selected & randomly selected \\ \hline
\end{tabular*}
\end{center}
\caption{The partitioning of the collected data into training and test
sets. The \digg{} data is divided by time while the \youtube{} videos
are chosen randomly for each set, respectively.}
\label{table:sets}
\end{table}

For reference time $t_r$ where we intend to predict the popularity of
submissions we chose 30 days after the submission time. Since the
predictions naturally depend on $t_i$ and how close we are to the
reference time, we performed the parameter estimations in hourly
intervals starting after the introduction of any submission.

Analogously to LSE and RSE, we will consider the following prediction
error measures for one particular submission $s$:
\begin{equation}
\mathrm{QSE}(s, t_i, t_r) = \left[ \hat N_s(t_i, t_r) - N_s(t_r)
  \right] ^ 2
\end{equation}
and
\begin{equation}
\mathrm{QRE}(s, t_i, t_r) = \left[ \frac{\hat N_s(t_i, t_r) -
    N_s(t_r)}{N_s(t_r)} \right] ^ 2.
\end{equation}
$\mathrm{QSE}(s, t_i, t_r)$ is the squared difference between the
prediction and the actual popularity for a particular submission $s$,
and QRE is the relative squared error. We will use this notation to
refer to their ensemble average values, too, $\mathrm{QSE} = \langle
\mathrm{QSE}(c, t_i, t_r) \rangle_c$, where $c$ goes over all
submissions in the test set, and similarly, $\mathrm{QRE} = \langle
\mathrm{QRE}(s, t_i, t_r) \rangle_c$. We used the parameters obtained
in the learning session to perform the predictions on the test set,
and plotted the resulting average error values calculated with the
above error measures.
Figure~\ref{figure:digg-youtube-prediction_quality} shows QSE and QRE
as a function of $t_i$, together with their respective standard
deviations. $t_i$, as earlier, is measured from the time a video is
presented in the recent list or when a story gets promoted to the
front page of \digg{}.

\newcommand{\thiswidth}{0.62\columnwidth}
\begin{figure*}
\begin{center}
(a)\includegraphics[width=\thiswidth]{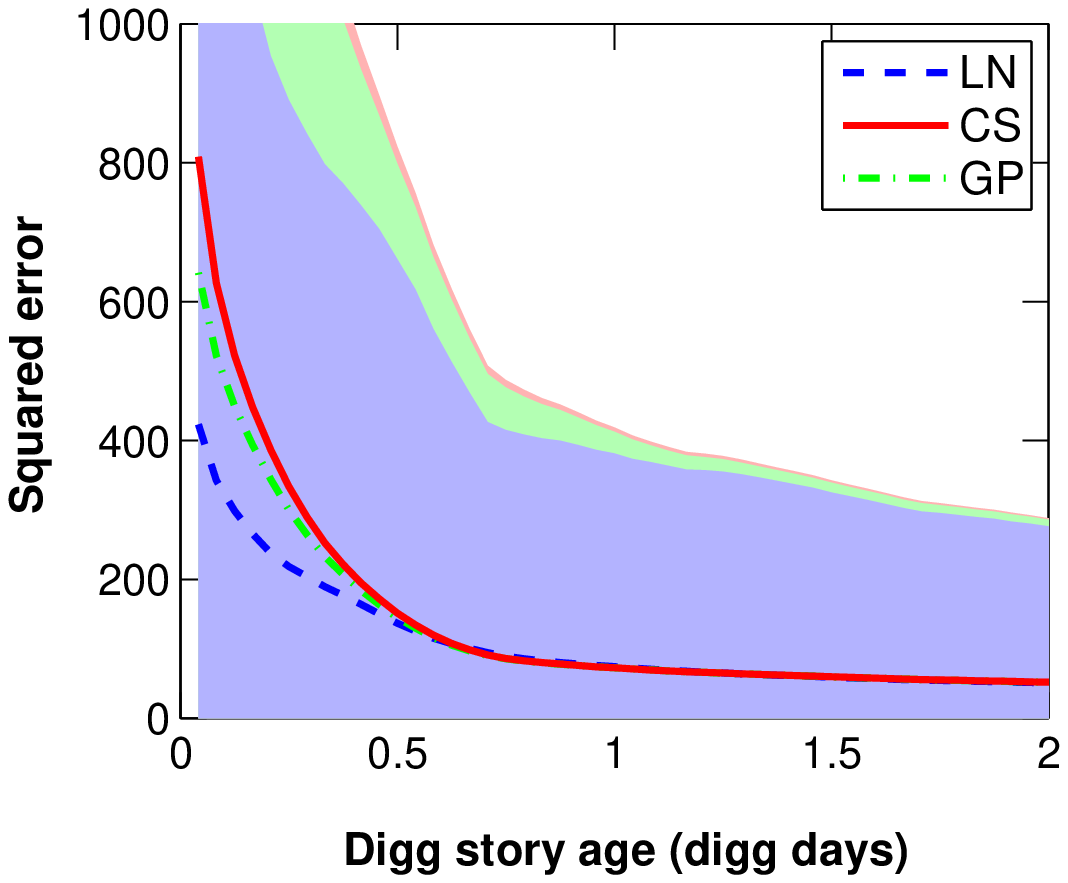}
\hspace{0.5cm}
(b)\includegraphics[width=\thiswidth]{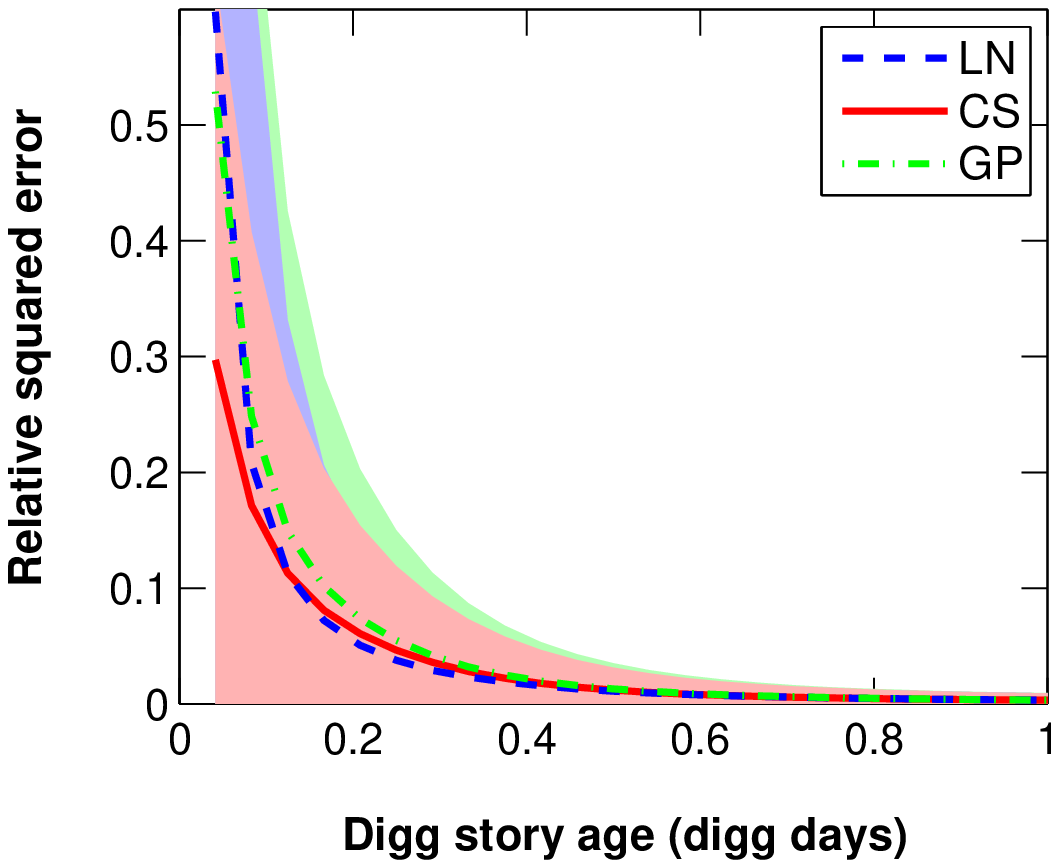}
\end{center}
\begin{center}
(c)\includegraphics[width=\thiswidth]{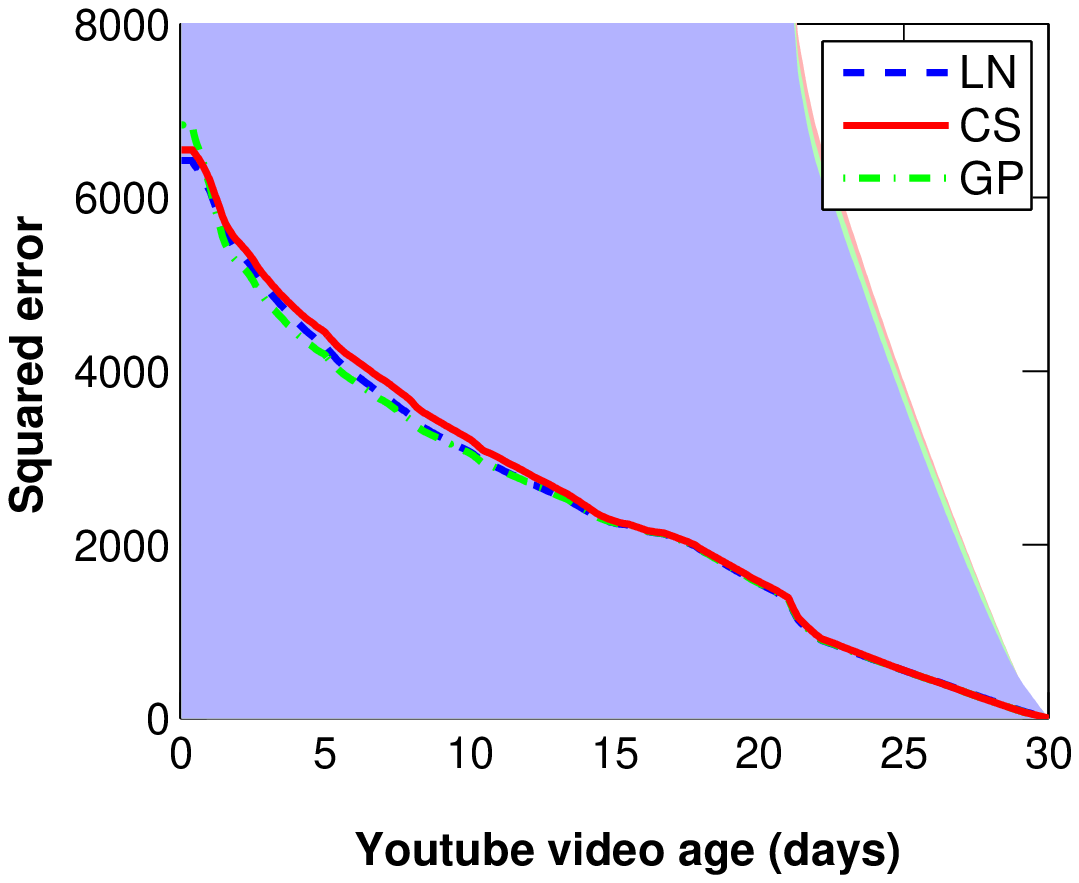}
\hspace{0.5cm}
(d)\includegraphics[width=\thiswidth]{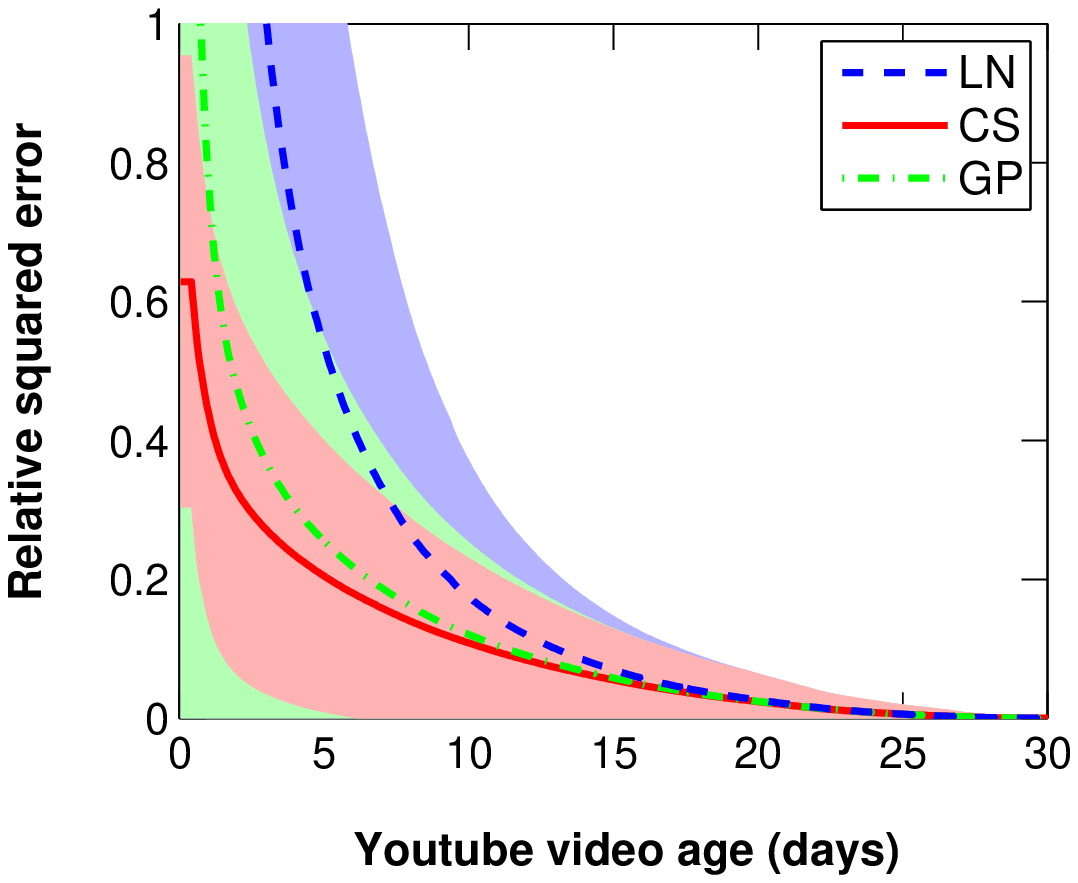}
\end{center}
\caption{The performance of the different prediction models, measured
  by two error functions as defined in the text: the absolute squared
  error QSE [\textbf{(a)} and \textbf{(c)}], and the relative squared
  error QRE [\textbf{(b)} and \textbf{(d)}], respectively.
  \textbf{(a)} and \textbf{(b)} show the results for \digg{}, while
  \textbf{(c)} and \textbf{(d)} for Youtube. The shaded areas indicate
  one standard deviation of the individual submission errors around
  the average.}
\label{figure:digg-youtube-prediction_quality}
\end{figure*}

QSE, the squared error is indeed smallest for the LN model for \digg{}
stories in the beginning, then the difference between the three models
becomes modest. This is expected since the LN model optimizes for the
RSE objective function, which is equivalent to QSE up to a constant
factor. \youtube{} videos do not show remarkable differences against
any of the three models, however. A further difference between \digg{}
and \youtube{} is that QSE shows considerable dispersion for
\youtube{} videos over the whole time axis, as can be seen from the
large values of the standard deviation (the shaded areas in
Fig.~\ref{figure:digg-youtube-prediction_quality}). This is
understandable, however, if we consider that the popularity of \digg{}
news saturates much earlier than that of \youtube{} videos, as will be
studied in more detail in the following section.

Considering further Fig.~\ref{figure:digg-youtube-prediction_quality}
(b) and (d), we can observe that the relative expected error QRE
decreases very rapidly for \digg{} (after 12 hours it is already
negligible), while the predictions converge slower to the actual value
in the case of \youtube{}. Here, however, the CS model outperforms the
other two for both portals, again as a consequence of fine-tuning the
model to minimize the objective function RSE. It is also apparent that
the variation of the prediction error among submissions is much
smaller than in the case of QSE, and the standard deviation of QRE is
approximately proportional to QRE itself. The explanation for this is
that the noise fluctuations around the expected average as described
by Eq.~(\ref{equation:linear_product}) are additive on a
\emph{logarithmic} scale, which means that taking the ratio of a
predicted and an actual popularity as in QRE is translated into a
difference on the logarithmic scale of popularities. The difference of
the logs is commensurate with the noise term in
Eq.~(\ref{equation:linear_product}), thus stays bounded in QRE, and is
instead amplified multiplicatively in QSE.

In conclusion, for relative error measures the CS model should be
chosen, while for absolute measures the LN model is a good choice.

\section{Saturation of the popularity}
\label{section:saturation}

\begin{figure}
\begin{center}
\includegraphics[width=\figwidth]{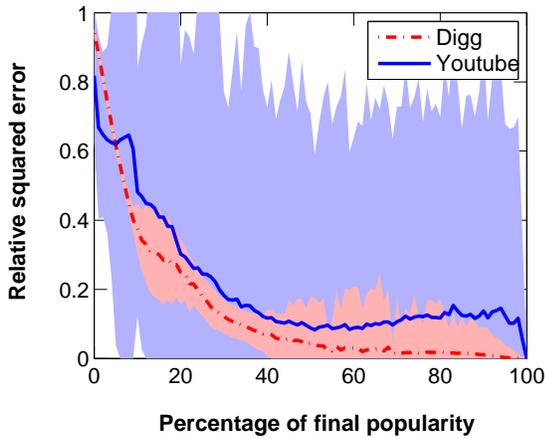}
\end{center}
\caption{The relative squared error shown as a function of the
  percentage of the final popularity of submissions on day 30. The
  standard deviations of the errors are indicated by the shaded
  areas.}
\label{figure:both-errors-pop}
\end{figure}

\begin{figure}
\begin{center}
\includegraphics[width=\figwidth]{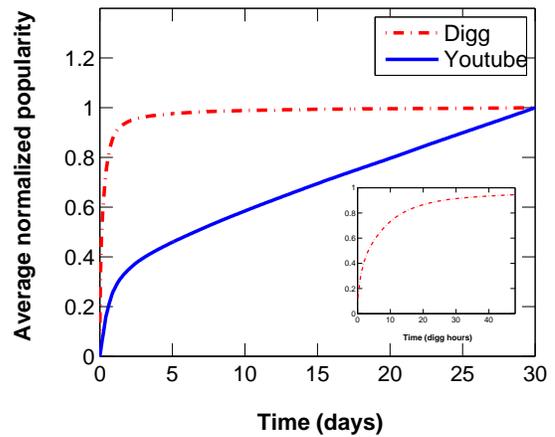}
\end{center}
\caption{Average normalized popularities of submissions for \youtube{}
  and \digg{} by the popularity at day 30. The inset shows the same
  for the first 48 digg hours of \digg{} submissions.}
\label{figure:both-average-normalized-pop}
\end{figure}

Here we discuss how the trends in the growth of popularities in time
are different for \youtube{} and \digg{}, and how this generally
affects the predictions.

As seen in the previous section, the predictions converge much faster
for \digg{} articles than for videos on \youtube{} to their respective
reference values, and the explanation can be found when we consider
how the popularity of submissions approaches the reference values. In
Fig.~\ref{figure:both-errors-pop} we show an analogous interpretation
of QRE, but instead of plotting the error against time, we plotted it
as a function of the actual popularity, expressed as the fraction of
reference value $N_s(t_r)$. The plots are averages over all content in
the test set, and over times $t_i$ in hourly increments up to $t_r$.
This means that the predictions across \youtube{} and \digg{} become
comparable, since we can eliminate the effect of the different time
dynamics imposed on content popularity by the visitors that are
idiosyncratic to the two different portals: the popularity of \digg{}
submissions initially grows much faster, but it quickly saturates to a
constant value, while \youtube{} videos keep getting views constantly
(Fig.~\ref{figure:both-average-normalized-pop}). As
Fig.~\ref{figure:both-errors-pop} shows, the average error QRE for
\digg{} articles converges to $0$ as we approach the reference time,
with variations in the error staying relatively small. On the other
hand, the same error measure does not decrease monotonically for
\youtube{} videos until very close to the reference, which means that
the growth of popularity of videos still shows considerable
fluctuations near the 30st day, too, when the popularity is already
almost as large as the reference value.

This fact is further illustrated by
Fig.~\ref{figure:both-average-normalized-pop}, where we show the
average normalized popularities for all submissions. This is
calculated by dividing the popularity counts of individual submission
by their reference popularities on day 30, and averaging the resulting
normalized functions over all content. An important difference that is
apparent in the figure is that while \digg{} stories saturate fairly
quickly (in about one day) to their respective reference popularities,
\youtube{} videos keep getting views \emph{all throughout their
  lifetime} (at least throughout the data collection period, but it is
expected that the trendline continues almost linearly). The rate at
which videos keep getting views may naturally differ among videos:
less popular videos in the beginning are likely to show a slow pace
over longer time scales, too. It is thus not surprising that the
fluctuations around the average are not getting supressed for videos
as they age (compare with Fig.~\ref{figure:both-errors-pop}). We also
note that the normalized growth curves shown in
Fig.~\ref{figure:both-average-normalized-pop} are exactly $P(t_i,
t_r)$ of Eq.~(\ref{equation:norm-popularity}) when $t_r = 30$ days.

The mechanism that gives rise to these two markedly different
behaviors is a consequence of the different ways of how users find
content on the two portals: on \digg{}, articles become obsolete
fairly quickly, since they oftenmost refer to breaking news, fleeting
Internet fads, or technology-related stories that naturally have a
limited time period while they interest people. Videos on \youtube{},
however, are mostly found through search, since due to the sheer
amount of videos uploaded constantly it is not possible to match
\digg{}'s way of giving exposure to each promoted story on a front
page (except for featured videos, but here we did not consider those
separately). The faster initial rise of the popularity of videos can
be explained by their exposure on the ``recently added'' tab of
\youtube{}, but after they leave that section of the site, the only
way to find them is through keyword search or when they are displayed
as related videos with another video that is being watched. It serves
thus an explanation to why the predictions converge faster for \digg{}
stories than \youtube{} videos (10\% accuracy is reached within about
2 hours on \digg{} vs. 10 days on \youtube{}) that the popularities of
\digg{} submissions do not change considerably after 2 days.

\section{Conclusions and related work}
\label{section:conclusions}
 
In this paper we presented a method and experimental verification on
how the popularity of (user contributed) content can be predicted very
soon after the submission has been made, by measuring the popularity
at an early time. A strong linear correlation was found between the
logarithmically transformed popularities at early and later times,
with the residual noise on this transformed scale being normally
distributed. Using the fact of linear correlation we presented three
models for making predictions about future popularity, and compared
their performance on \youtube{} videos and \digg{} story submissions.
The multiplicative nature of the noise term allows us to show that the
accuracy of the predictions will exhibit a large dispersion around the
average if a direct squared error measure is chosen, while if we take
the relative errors the dispersion is considerably smaller. An
important consequence is that absolute error measures should be
avoided in favor of relative measures in community portals when the
error of the prediction is estimated.

We mention two scenarios where predictions of individual content can
be used: advertising and content ranking. If the popularity count is
tied to advertising revenue such as what results from advertisement
impressions shown beside a video, the revenue may be fairly accurately
estimated, since the uncertainty of the relative errors stays
acceptable. However, when the popularities of different content are
compared to each other as commonly done in ranking and presenting the
most popular content to users, it is expected that the precise
forecast of the ordering of the top items will be more difficult due
to the large dispersion of the popularity count errors.

We based the predictions of future popularities only on values
measurable in the present, but did not consider the semantics of
popularity and \emph{why} some submissions become more popular than
others. We believe that in the presence of a large user base
predictions can essentially be made on observed early time series, and
semantic analysis of content is more useful when no early clickthrough
information is known for content. Furthermore, we argue for the
generality of performing maximum likelihood estimates for the model
parameters in light of a large amount of experimental information,
since in this case Bayesian inference and maximum likelihood methods
essentially yield the same estimates~\cite{higgins:1977}.

There are several areas that we could not explore here. It would be
interesting to extend the analysis by focusing on different sections
of the Web 2.0 portals, such as how the ``news \& politics'' category
differs from the ``entertainment'' section on \youtube{}, since we
expect that news videos reach obsolescence sooner than videos that are
recurringly searched for for a long time. It is also to be seen if it
is possible to forecast a \digg{} submission's popularity when the
diggs are coming from a small number of users only whose voting
history is known, as is the case for stories in the upcoming section
of \digg{}.

In related works video on demand systems and properties of media files
on the Web have been studied in detail, statistically characterizing
video content in terms of length, rank, and
comments~\cite{cheng:2007,acharya00characterizing,1111629}. Video
characteristics and user access frequencies are studied together when
streaming media workload is
estimated~\cite{1298310,1251441,1242804,1217968}. User participation
and content rating is also modeled in \digg{}, with particular
emphasis on the social network and the upcoming phase of
stories~\cite{Lerman07ic}. Activity fluctuations, user commenting
behavior prediction, the ensuing social network, and community
moderation structure is the focus of studies on
\slashdot{}~\cite{1318432,citeulike:3041174,985761}, a portal that is
similar in spirit to \digg{}. The prediction of user clickthrough
rates as a function of document and search engine result ranking order
has overlaps with this paper~\cite{1242642,1148175}. While the display
ordering of submissions plays a less important role for the
predictions presented here, Dupret et al. studied the effect of
document position in a list on its selection probability with a
Bayesian network model that becomes important when static content is
predicted~\cite{DBLP:conf/spire/DupretPHM06}; a related area is online
ad clickthrough rate prediction also~\cite{1242643}.

%\section{Acknowledgments}

%We are thankful to Fang Wu, Tad Hogg, and Dennis Wilkinson for
%discussions about this work, and to Simla Ceyhan for collecting the
%data on \youtube{} video views.

%\bibliographystyle{abbrv}
\bibliographystyle{habbrv}
\bibliography{paper}

\end{document}